\documentclass[preprint,11pt]{elsarticle}

\usepackage{fancyhdr}
\usepackage{fullpage}
\usepackage{amsmath, mathtools}
\usepackage{amsbsy}
\usepackage{amssymb}
\usepackage{amscd}
\usepackage{amsfonts}
\usepackage{supertabular}
\usepackage{tablefootnote}
\usepackage{graphics}
\usepackage{verbatim}
\usepackage{epsfig}
\usepackage{xspace}
\usepackage{euscript}
\usepackage{alltt}
\usepackage{boxedminipage}
\usepackage{float}
\usepackage[colorlinks]{hyperref}
\usepackage{color}
\usepackage[all]{xy}
\usepackage{t1enc}
\usepackage{times,exscale}
\usepackage{graphicx,calc}
\usepackage{subfig}
\usepackage[ruled,vlined]{algorithm2e}
\biboptions{sort&compress}
\usepackage{epstopdf}
\usepackage{array,multirow}

\usepackage{graphicx}					
\usepackage{amsmath,amssymb}		
\usepackage{subfig}					
\usepackage{multirow}
\usepackage{booktabs}

\usepackage{comment}

\newcommand{\eminus}[1]{\ensuremath{{\times}10^{-#1}}}

\journal{arXiv}
\begin{document}

	\begin{frontmatter}
		
		\title{Spectral scheme for atomic structure calculations in density functional theory}
		\author[gatech_coe]{Sayan Bhowmik}
            \author[llnl]{John E. Pask}
		\author[gatech_coe]{Andrew J. Medford}
            \author[gatech_coe,gatech_coc]{Phanish Suryanarayana\corref{cor}}
            
		\address[gatech_coe]{College of Engineering, Georgia Institute of Technology, Atlanta, GA 30332, USA}
        \address[gatech_coc]{College of Computing, Georgia Institute of Technology, Atlanta, GA 30332, USA}
        \address[llnl]{Physics Division, Lawrence Livermore National Laboratory, Livermore, CA 94550, USA}
		
		\cortext[cor]{Corresponding Author (\it phanish.suryanarayana@ce.gatech.edu) }
		
		\begin{abstract}
         We present a spectral scheme for atomic structure calculations in pseudopotential Kohn-Sham density functional theory. In particular, after applying an exponential transformation of the radial coordinates, we employ global polynomial interpolation on  a Chebyshev grid, with derivative operators approximated using the  Chebyshev differentiation matrix, and integrations using  Clenshaw-Curtis quadrature. We demonstrate the accuracy and efficiency of the scheme through spin-polarized and unpolarized calculations for representative atoms, while considering  local, semilocal, and hybrid exchange-correlation functionals. In particular, we find that $\mathcal{O}$(200) grid points are sufficient to achieve an accuracy of 1 microhartree in the eigenvalues  for optimized norm conserving Vanderbilt pseudopotentials spanning the periodic table from atomic number $Z = 1$ to $83$.
		\end{abstract}
		
		
	\end{frontmatter}

	\section{Introduction}
	Over the past few decades, Kohn-Sham Density Functional Theory (DFT) \cite{kohn1965self, hohenberg1964inhomogeneous} has  established itself as  a cornerstone of the materials and chemical sciences research, providing an avenue to understand and predict material and chemical  properties from the first principles of quantum mechanics. The widespread use of DFT can be attributed to not only its high accuracy-to-cost ratio compared to other such ab initio methods, but also to its generality, being applicable to  isolated systems such as molecules and clusters, semi-infinite systems such as  nanotubes and surfaces, as well as bulk 3D systems, at both ambient conditions \cite{burke2012perspective} and extreme conditions of temperature and pressure \cite{stanek2024review}.

Despite significant advances, particularly in the context of large-scale parallelization \cite{gavini2022roadmap}, these calculations remain computationally expensive, which has motivated the development of numerous strategies to reduce their cost. These include exploiting the spatial symmetry in the system to reduce the computations to the unit cell or fundamental domain, e.g., translational symmetry of a crystal \cite{martin2020electronic}, cyclic symmetry observed in certain molecules/clusters and 1D nanostructures \cite{ghosh2019symmetry, banerjee2016cyclic}, helical symmetry found in certain 1D nanostructures  \cite{sharma2021real, bhardwaj2024strain}, and spherical symmetry of isolated atoms \cite{hamann2013optimized, fuchs1999ab}.

Atomic structure calculations \cite{lehtola2019review}, wherein the electronic structure problem is solved in radial coordinates for isolated atoms, form an integral part of the DFT  infrastructure.  In particular, they are used for inverting  the radial Kohn-Sham equations during generation of pseudopotentials \cite{hamann2013optimized, troullier1991efficient, shojaei2023soft}, which are critical for efficient and accurate DFT calculations. In addition, they are used for generating an atom-centered basis in the projector augmented wave (PAW)  \cite{blochl1994projector, holzwarth2001projector} method, which  allows for coarser grid spacings relative to norm conserving pseudopotentials,  and also in Hubbard-corrected DFT (DFT+U) \cite{anisimov1991density, anisimov1991band}, which allows for the study of strongly correlated systems.  The  densities, orbitals, and potentials from atomic structure calculations are used to generate  initial guesses for accelerating the convergence of DFT calculations \cite{lehtola2019assessment, lehtola2020efficient, van2006starting, ghosh2017sparcI, ghosh2017sparc2}. In addition, the orbitals form a part of the numerical atomic orbitals (NAO) scheme \cite{CONQUEST, garcia2020siesta, GPAW, blum2009ab} within the linear combination of atomic orbitals (LCAO) method, and the densities form a part of the Gaussian multipole (GMP) scheme \cite{lei2022universal, timmerman2024overcoming} for generating machine-learned force fields (MLFFs). Isolated atom orbitals also facilitate the analysis of the results of DFT simulations, e.g.,  projected density of states (PDOS) \cite{herath2020pyprocar, wang2021vaspkit, ward2018matminer}.

A number of solution strategies have been developed for atomic structure calculations, including shooting-type methods \cite{hamann2013optimized, holzwarth2001projector, kotochigova1997local, desclaux1971relativistic, jonsson2007grasp2k, fuchs1999ab, vcertik2013dftatom, OLIVEIRA2008524}, Numerov scheme \cite{tobin1975eigenvalue, biegler1986nonrelativistic, andrae1997numerical, andrae2000numerical}, spline approaches \cite{fischer2009b, grant2009b, holzwarth2022cubic}, finite-difference method \cite{cayford1974finite, AndraeExamination2001}, and finite element method \cite{lehtola2023meta, vcertik2024high, lehtola2020fully, lehtola2019fully, romanowski2007b, romanowski2009adaptive, OZAKI20111245}. However, many of these methods focus on solving the non-interacting Schr\"odinger and/or Dirac equations, while only a limited number are specifically designed for Kohn-Sham DFT \cite{hamann2013optimized, holzwarth2001projector, vcertik2013dftatom, holzwarth2022cubic, lehtola2023meta, vcertik2024high, lehtola2020fully, fuchs1999ab, lehtola2019fully, romanowski2007b, romanowski2009adaptive, AndraeExamination2001, OLIVEIRA2008524}. Among these, most solvers are aimed at all-electron calculations, while only a few can incorporate the pseudopotential approximation \cite{hamann2013optimized, fuchs1999ab, OLIVEIRA2008524}, a common strategy to reduce the computational cost of DFT calculations. Even these are restricted to local exchange-correlation functionals, with semilocal and hybrid exchange-correlation functionals not implemented heretofore. Indeed, an atomic solver that can accommodate local, semilocal, and hybrid exchange-correlation functionals will enable improved accuracy and efficiency in the aforementioned applications of atomic structure calculations within the pseudopotential approximation. 

In this work, we present a spectral scheme for atomic structure calculations in pseudopotential Kohn-Sham DFT. In particular, after transforming the radial coordinates through an exponential mapping,  we employ polynomial interpolation on  a Chebyshev grid, on which derivative operators are approximated using the  Chebyshev differentiation matrix, and integrals are evaluated using  Clenshaw-Curtis quadrature. We demonstrate the accuracy and efficiency of the resulting scheme through spin-polarized and unpolarized calculations for representative atoms, while considering  local, semilocal, and hybrid exchange-correlation functionals. In particular, for the choice of optimized norm conserving Vanderbilt (ONCV) pseudopotentials \cite{hamann2013optimized} spanning the periodic table from atomic number $Z = 1$ to $83$,  we find that $\mathcal{O}$(200) grid points is sufficient to  achieve an accuracy of 1 microhartree or better in the eigenvalues.

The remainder of this manuscript is organized as follows. In Section~\ref{Sec:Equations}, we provide the mathematical background for atomic structure calculations. In Section~\ref{formulation}, we describe the spectral scheme, whose accuracy and efficiency is verified through representative examples in Section~\ref{results}. Finally, we provide concluding remarks in Section~\ref{conclusions}.

	\section{Atomic structure problem} \label{Sec:Equations}
The energy functional for an isolated atom in pseudopotential Kohn-Sham DFT takes the form:
\begin{equation} \label{Eq:EnergyFunctional}
\mathcal{E}[\widetilde{R}_{nl}^\sigma, \widetilde{V}_H] = T_{s}[\widetilde{R}_{nl}^\sigma] + E_{xc}[\rho^\sigma, |\nabla \rho^\sigma|, \tau^\sigma, \widetilde{R}_{nl}^\sigma] + K[\widetilde{R}_{nl}^\sigma] + E_{el}[\rho, \widetilde{V}_H] \,,
\end{equation}
where $\sigma \in \{\uparrow, \downarrow \}$ denotes the spin, with $\uparrow$ and $\downarrow$ denoting the spin-up and spin-down components, respectively; $n$, $l$, and $m$ are the principal, azimuthal, and magnetic quantum numbers, respectively, used to index the quantities; $R_{nl}^\sigma = \widetilde{R}_{nl}^{\sigma}/r$ is the radial part of the Kohn-Sham orbitals $\psi^{\sigma}_{nlm}$, i.e.,  $\psi^{\sigma}_{nlm}(\mathbf{x}) = R^{\sigma}_{nl}(r)Y_{lm}(\theta,\phi)$, with $Y_{lm}$ denoting the spherical harmonics and ($r$,$\theta$,$\phi$) representing the coordinates of the position vector $\mathbf{x}$ in the spherical polar coordinate system; $V_H= \widetilde{V}_H/r$ is the Hartree potential;  $\rho$ is the electron density, which is itself dependent on the orbitals:
        \begin{align} \label{density}
                \rho & = \sum_{\sigma} \rho^{\sigma}  = \frac{1}{4\pi r^2}\sum_{\sigma} \sum_{nl} g^\sigma_{nl} (\widetilde{R}^{\sigma}_{nl})^2 \,,
        \end{align}
with $g^\sigma_{nl}$ representing the angular momentum summed  occupations, i.e., $g^\sigma_{nl} = (2l+1) f^{\sigma}_{nlm}$, $f^{\sigma}_{nlm} \in \{0,1\}$ being the occupation of the Kohn-Sham orbital $\psi^{\sigma}_{nlm}$;   
\begin{equation} \label{Eq:derrhosigma}
|\nabla \rho^\sigma| = \left| \frac{d \rho^{\sigma}}{dr} \right| \,;
\end{equation}
$\tau^{\sigma}$ is the positive definite form of the kinetic energy density:
        \begin{equation}\label{tau}
                \tau^\sigma = \frac{1}{2}\sum_{nl}\frac{g^\sigma_{nl}}{4\pi}\left[ \left( \frac{1}{r}\frac{d \widetilde{R}^\sigma_{nl}}{dr} - \frac{\widetilde{R}^\sigma_{nl}}{r^2}\right)^2 + \frac{l(l+1)}{r^4}(\widetilde{R}^\sigma_{nl})^2\right] \,;
        \end{equation}   
 $E_{xc}$ is the exchange-correlation energy, for which different approximations exist; and $T_s$, $K$, and $E_{el}$ are the electronic kinetic energy,  nonlocal pseudopotential energy within the Kleinman-Bylander representation \cite{kleinman1980relativistic}, and electrostatic energy, respectively:
\begin{subequations}
\begin{align}
                T_{s}[\widetilde{R}_{nl}^\sigma] & = - \frac{1}{2} \sum_{\sigma} \sum_{nl} g^\sigma_{nl} \int  \left( \widetilde{R}^{\sigma}_{nl} (r) \frac{d^2 \widetilde{R}^\sigma_{nl} (r)}{dr^2} +  \widetilde{R}^{\sigma}_{nl}(r) \frac{l(l+1)}{r^2} \widetilde{R}^\sigma_{nl}(r) \right)  \, dr \,, \label{KE_eq} \\
K[\widetilde{R}_{nl}^\sigma] & = \sum_\sigma \sum_{nl} g^\sigma_{nl} \sum_{l} \gamma_{l} \left( \int  \widetilde{\chi}_l(r) \widetilde{R}^\sigma_{nl}(r) \, dr \right)^2 \,, \label{PSP_En_rad}    \\
E_{el}[\rho, \widetilde{V}_H] & = \max_{\widetilde{V}_H} \left( -\frac{1}{2} \int \left( \frac{d \widetilde{V}_H(r)}{d r} \right)^2 \, dr  + 4 \pi \int \rho(r) \widetilde{V}_{H}(r) r \, d r \right) + 4 \pi \int \rho(r) V_{loc}(r) r^2 \, d r \,, \label{ElecEnergy}
\end{align}
\end{subequations} 
with  $\chi_l = \widetilde{\chi}_{l}/r$ denoting the nonlocal pseudopotential projectors, $V_{loc}$ representing the local part of the pseudopotential, and the variational problem for $\widetilde{V}_H$ derived as in previous work \cite{pask2005real, Phanish2010, Phanish2011}. In obtaining the above expressions from the full 3D Kohn-Sham problem, the following relations have been used:
\begin{equation} \label{UnsoldThm}
\sum_{m} (Y_{lm})^2 = \frac{(2l+1)}{4\pi} \,, \quad \nabla^2 Y_{lm} = -\frac{l(l+1)}{r^2} Y_{lm} \,.
\end{equation}
Note that unless specified, the limits of integration for all integrals are zero and infinity.       
        
The electronic ground state is determined by the following variational problem:
\begin{align}
\min_{\widetilde{R}_{nl}^\sigma} \max_{\widetilde{V}_H} \mathcal{E}[\widetilde{R}_{nl}^\sigma,  \widetilde{V}_H] \quad  \text{s.t.} \quad \int \ \widetilde{R}^{\sigma}_{nl}(r) \widetilde{R}^{\sigma}_{n'l'}(r) \, dr  = \delta_{nn'} \delta_{l l'} \,,
\end{align}         
where the constraint arises from the requirement of orthonormality between the Kohn-Sham orbitals. The corresponding Euler-Lagrange equations take the form:
        \begin{subequations}\label{ground_state}
            \begin{align}
                 \left[ \mathcal{H}\equiv -\frac{1}{2}\frac{d^2}{dr^2} + V_{xc} + V_{nl} + \frac{\widetilde{V}_H}{r} + V_{loc} + \frac{l(l+1)}{2 r^2} \right] \widetilde{R}^\sigma_{nl} = \lambda_{nl}\widetilde{R}^\sigma_{nl} \,, \nonumber \\
                 \hspace{57mm} \widetilde{R}^\sigma_{nl}(r=0) = 0 \,, \widetilde{R}^\sigma_{nl}(r \rightarrow \infty) = 0 \,; \label{EL:Eigenproblem} \\
                 \frac{d^2 \widetilde{V}_H(r)}{dr^2} = -4\pi r \rho(r) \,, \quad  \widetilde{V}_H(r =0) =0 \,,  \widetilde{V}_H(r \rightarrow \infty) = Z \,; \label{EL:Poisson}
            \end{align}
        \end{subequations}
where $\mathcal{H}$ represents the Hamiltonian, $Z$ is the number of electrons in valence, $V_{nl}$ is the nonlocal pseudopotential operator:
        \begin{equation}\label{Vnl}
            V_{nl}\widetilde{R}^\sigma_{nl} = \sum_l \gamma_{l}\widetilde{\chi}_l \int \widetilde{\chi}_l(r') \widetilde{R}^\sigma_{nl}(r') \, dr' \,, 
        \end{equation}
and $V_{xc}$ is the exchange-correlation potential, obtained by taking variation of the exchange-correlation energy $E_{xc}$. The exchange-correlation energy takes the following form for local and semilocal exchange-correlation functionals:
        \begin{equation}\label{Exc}
                E_{xc}(\rho^\sigma, |\nabla \rho^\sigma|, \tau^\sigma) = 4\pi \int  \rho(r)\varepsilon_{xc}(\rho^\sigma, |\nabla \rho^\sigma|, \tau^\sigma)\, r^2 dr \,,
        \end{equation}
whereby the exchange-correlation potential takes the form: 
\begin{subequations}
\begin{align}
V_{xc}^{LDA} \widetilde{R}^\sigma_{nl} & =  \left[  \varepsilon_{xc} + \rho^\sigma\frac{\partial \varepsilon_{xc}}{\partial \rho^\sigma}  \right]   \widetilde{R}^\sigma_{nl} \,, \\
V_{xc}^{GGA} \widetilde{R}^\sigma_{nl} &= V_{xc}^{LDA} \widetilde{R}^\sigma_{nl}  + \left[ - \frac{d}{dr}\left( \rho^\sigma\frac{\partial \varepsilon_{xc}}{\partial |\nabla \rho^\sigma|}\right) - \frac{2\rho^\sigma}{r}\frac{\partial \varepsilon_{xc}}{\partial |\nabla \rho^\sigma|} \right] \widetilde{R}^\sigma_{nl} \,, \\
V_{xc}^{mGGA} \widetilde{R}^\sigma_{nl} &= V_{xc}^{GGA} \widetilde{R}^\sigma_{nl} + \frac{1}{2}\left[  \frac{\partial \varepsilon_{xc}}{\partial \tau^\sigma}\frac{l(l+1)}{r} \widetilde{R}^\sigma_{nl} -  \frac{d}{dr}\left( r \frac{\partial \varepsilon_{xc}}{\partial \tau^\sigma} \frac{d\widetilde{R}^\sigma_{nl}}{dr}\right) - \frac{d}{dr} \left( \frac{\partial \varepsilon_{xc}}{\partial \tau^\sigma}\widetilde{R}^\sigma_{nl} \right)  \right] \,,
\end{align}
\end{subequations}
with LDA denoting the local density approximation, GGA denoting the generalized gradient approximation, and mGGA denoting the meta-GGA approximation. In the  case of hybrid exchange-correlation functionals, the exchange-correlation energy incorporates a certain fraction of the exact exchange energy, which can be reduced to the following form for isolated atoms:
\begin{align}
           E_X[\widetilde{R}_{nl}^\sigma]   = -\frac{1}{2} \sum_{\substack{\sigma \\ \sigma^\prime}} \sum_{\substack{nl \\ n'l'}} g^\sigma_{nl} g^{\sigma^\prime}_{n'l'} \sum_{l^{\prime \prime}=|l-l'|}^{l+l'} \left[ 
                \begin{pmatrix}
                    l & l' & l^{\prime \prime} \\
                    0 & 0 & 0
                \end{pmatrix}^2 
                \int  \widetilde{R}^\sigma_{nl}(r) \widetilde{R}^{\sigma^\prime}_{n'l'}(r) \, dr 
                \int  \frac{r^{l^{\prime \prime}}_{<}}{r^{l^{\prime \prime}+1}_{>}} \widetilde{R}^\sigma_{nl}(r') \widetilde{R}^{\sigma^\prime}_{n'l'}(r') \, dr' \right] \,,
\end{align}
where $\begin{pmatrix} l & l' & l^{\prime \prime} \\ 0 & 0 & 0 \end{pmatrix}$ is the Wigner-3j symbol used for adding angular momenta, $r_<$ is $\min(r,r')$ and $r_>$ is $\max(r,r')$. The corresponding potential can be written as \cite{cinal2020highly}:
\begin{equation}\label{Vxx}
            V_X \widetilde{R}_{nl}^\sigma = -\frac{1}{2} \sum_{\substack{\sigma^\prime \\ n'l'}} g^{\sigma^\prime}_{n'l'} \sum_{l^{\prime \prime}=|l-l'|}^{l+l'} \left[
            \begin{pmatrix}
                    l & l' & l^{\prime \prime} \\
                    0 & 0 & 0
            \end{pmatrix}^2
            \widetilde{R}^{\sigma^\prime}_{n'l'} \int  \frac{r^{l^{\prime \prime}}_{<}}{r^{l^{\prime \prime}+1}_{>}} \widetilde{R}^{\sigma^\prime}_{n'l'}(r') \widetilde{R}_{nl}^\sigma(r') \, dr' \right] \,,
        \end{equation}
which now forms a part of the Hamiltonian. 

\section{Spectral scheme}  \label{formulation}
In this section, we present a spectral scheme for solution of the atomic structure problem in Kohn-Sham DFT. In particular, given the Dirichlet boundary conditions on both $\widetilde{R}^\sigma_{nl}$ and $\widetilde{V}_H$,  we employ algebraic polynomial interpolation. Since the polynomials used for interpolation are global in nature, the resulting scheme inherits the features of a spectral scheme, and hence is referred to as such here \cite{trefethen2000spectral}.  
	
The Kohn-Sham orbitals are expected to be more oscillatory near the nucleus, while smoothly decaying to zero at larger distances. In view of this, we restrict the maximum radial coordinate to $r = R$, i.e.,  $r \in [0, R]$, and employ the following transformation:
\begin{align}
y = 2 \left(\frac{1-e^{\beta r}}{1 - e^{\beta R}} \right) - 1 \,,  \label{Eqn:transformation}
\end{align}
where the parameter $\beta<0$. This transformation  is analogous to the exponential/logarithmic grid used in atomic structure calculations \cite{vcertik2024high, vcertik2013dftatom}, along with scaling and shifting to ensure  $y \in [-1,1]$. 
Using the chain rule, the derivatives appearing in Eqn.~\ref{ground_state} can then be transformed as:
\begin{subequations}
\begin{align}
\frac{d}{dr} & := -\frac{2 \beta}{1 - e^{\beta R}}  \left[ 1 - \frac{1}{2}(y+1)(1-e^{\beta R}) \right] \frac{d}{dy} \,, \label{Eq:DerTrans} \\
\frac{d^2}{dr^2} & := -\frac{2 \beta^2}{1 - e^{\beta R}}  \left[ 1 - \frac{1}{2}(y+1)(1-e^{\beta R}) \right] \frac{d}{dy} + \left( -\frac{2 \beta}{1 - e^{\beta R}}  \left[ 1 - \frac{1}{2}(y+1)(1-e^{\beta R}) \right] \right)^2 \frac{d^2}{dy^2} \,.
\end{align}
\end{subequations}
The integrals appearing in Eqn.~\ref{ground_state}  and Eqn.~\ref{Eq:EnergyFunctional} can be transformed as:
\begin{align}
\int_{0}^R f(r) \, dr = \int_{-1}^1 f \left( \frac{1}{\beta} \log \left( 1 - \frac{1}{2} (y+1)(1-e^{\beta R})\right) \right) \left(  \frac{-\frac{1}{2}(1-e^{\beta R})}{\beta (1-\frac{1}{2}(y+1)(1-e^{\beta R}))}   \right) \, dy \,,
\end{align}
where $f$ is any given function and the second term is the Jacobian of the transformation. The boundary conditions on $\widetilde{R}^\sigma_{nl}$ and $\widetilde{V}_{H}$ in the transformed coordinates take the form:
\begin{subequations}
\begin{align}
\widetilde{R}^\sigma_{nl}(y=-1) = 0 \,,  \quad \widetilde{R}^\sigma_{nl}(y = 1) = 0 \,, \\
\widetilde{V}_{H}(y=-1) = 0 \,, \quad \widetilde{V}_H (y = 1) = Z \,.
\end{align}
\end{subequations}

        \begin{figure}[htbp]
            \centering
            \includegraphics[width=0.50\textwidth]{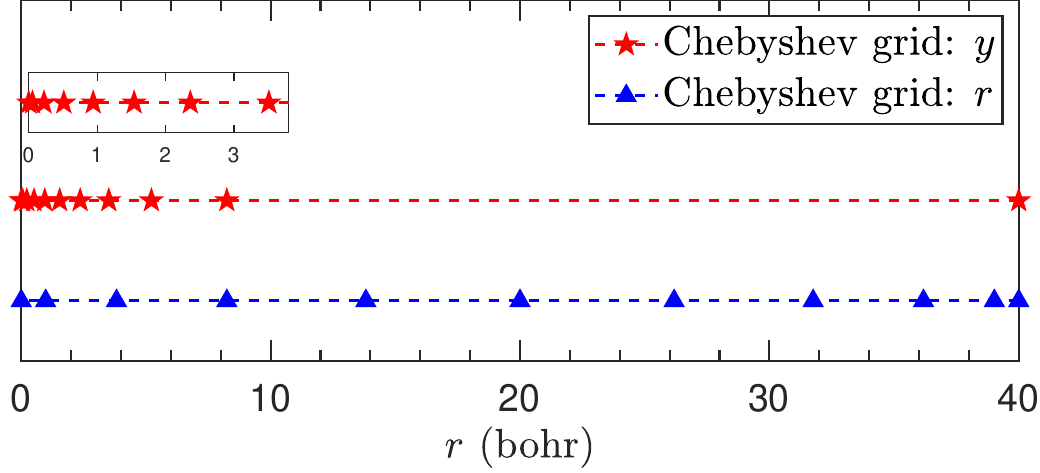}
            \caption{Grid point locations for the Chebyshev grids in $r$ and $y$ ($\beta=-0.45$) for $N_d=10$.}
            \label{fig:grid}
        \end{figure}
	
\begin{figure}[htbp]
            \centering
            \includegraphics[width=0.5\textwidth]{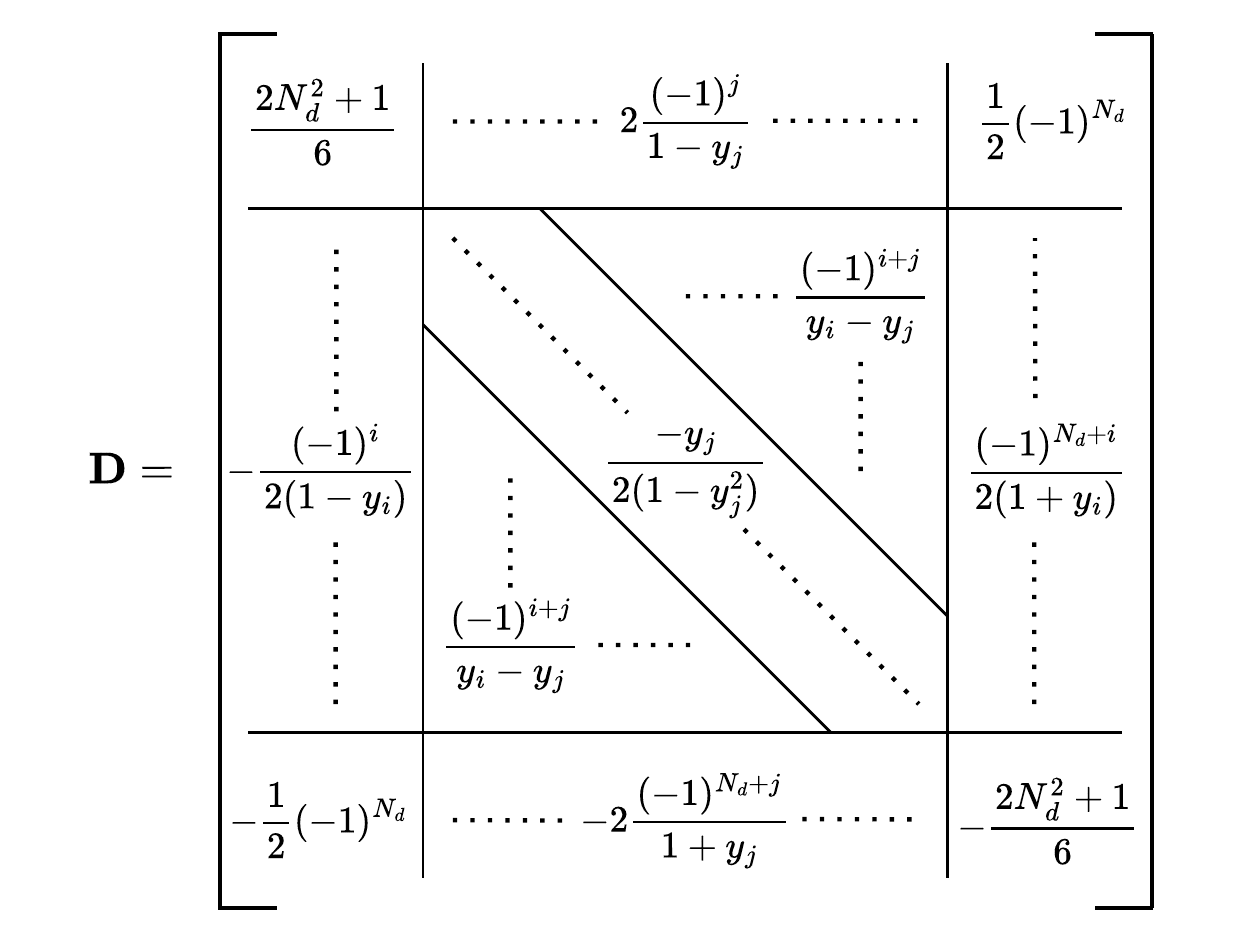}
            \caption{Pictorial representation of the Chebyshev differentiation matrix.}
            \label{fig:Dmatrix}
        \end{figure}

We employ a Chebyshev grid for discretization:
        \begin{equation}\label{ChebGrid}
            y_j = \cos{(j\pi/N_d)},\ \ \ j = 0, 1, \cdots, N_d \,,
        \end{equation}
which results in grid points as shown in Fig.~\ref{fig:grid}. Indeed, the transformation used in Eqn.~\ref{Eqn:transformation} ensures that the resulting grid is concentrated near the nucleus with decreasing resolution away from it, consistent with the physics of the  problem. The  corresponding Chebyshev differentiation matrix $\mathbf{D}$, which represents the discrete form of the $d/dy$ operator, takes the form \cite{trefethen2000spectral}:
        \begin{equation}\label{ChebDMat}
            [\mathbf{D}]_{ij} = 
            \begin{dcases}
             \frac{2N_d^2+1}{6} & i=j=0 \,,  \\
                -\frac{y_j}{2(1-y_j^2)} & i=j=1,2,\cdots,N_d-1 \,, \\
                               \frac{c_i}{c_j}\frac{(-1)^{i+j}}{(y_i-y_j)} & i \neq j,\ i,j=1,2,\cdots,N_d-1,\ c_k =  
                \begin{cases} 
                    2 & k \in\{0, N_d\} \,, \\
                    1 & \text{otherwise} \,,
                \end{cases} \\
                -\frac{2N_d^2+1}{6} & i=j=N_d \,,
            \end{dcases}
        \end{equation}
whose pictorial representation can be found in Fig.~\ref{fig:Dmatrix}. The discrete form of the $d^2/dy^2$ operator is then  given by $\mathbf{D}^2$. The zero Dirichlet boundary conditions on $\widetilde{R}^\sigma_{nl}$ at $y \in \{ -1, 1 \}$ are enforced by removing the first and last rows/columns of  $\mathbf{D}$ and $\mathbf{D}^2$ while solving the eigenproblem in  Eqn.~\ref{EL:Eigenproblem}. The zero Dirichlet boundary condition on $\widetilde{V}_H$ at $y=-1$  is enforced by removing the last row and column from 
$\mathbf{D}$ and $\mathbf{D}^2$, while the nonzero Dirichlet boundary condition at $y=1$ is enforced by replacing the first row of the matrix $\mathbf{L}$ resulting from the Poisson equation (Eqn.~\ref{EL:Poisson}) as: $[\mathbf{L}]_{00}=1$, $[\mathbf{L}]_{0j}=0$, $j=1, 2 \ldots, N_d$. Note that for the evaluation of the potential and energy in GGA, meta-GGA, and hybrid exchange-correlation functionals, $d \rho^{\sigma}/dr$ and therefore $d \rho^{\sigma}/dy$ needs to be evaluated. Since there is no particular boundary condition associated with $\rho^{\sigma}$  at $y=-1$, $\mathbf{D}$ cannot directly be used, hence we use the following relation:
\begin{equation}
            \frac{d\rho^{\sigma}}{dr} = \frac{1}{r}\left[ \frac{d(r \rho^\sigma)}{dr} - r \right] \,,
\end{equation}
with the derivative then transformed as described by Eqn.~\ref{Eq:DerTrans}. In particular, $r \rho^\sigma$ satisfies zero Dirichlet boundary conditions at $y \in \{ -1, 1 \}$, whereby the derivative operator $d/dy$ can be approximated by $\mathbf{D}$, with the first and last rows and columns removed, as described above. The integrals are evaluated using Clenshaw-Curtis quadrature \cite{trefethen2000spectral}:
\begin{align}
\int_{-1}^1 f(y) \, dy \approx \sum_{j=0}^{N_d} w_j f(y_j) \,,
\end{align}
where $y_j$ and $w_j$ are the nodes and weights of the quadrature rule, and $f$ is any given function. Indeed, the nodes of the Clenshaw-Curtis  quadrature coincide with the Chebyshev grid used for the discretization, which allows for highly efficient evaluation of the integrals.  Since we are using the same grid for interpolation and quadrature, the quadrature rule is exact only for polynomials of degree $N_d$. Therefore, the integrals are not evaluated exactly, whereby the overall scheme is strictly pseudospectral rather than spectral. However, we have found that the quadrature error is typically orders of magnitude smaller than the interpolation error, whereby the scheme retains spectral convergence.

	\section{Results and discussion}  \label{results}
We have implemented the spectral scheme described in the previous section in M-SPARC \cite{xu2020m, zhang2023version}, which is a \textsc{MATLAB} version of the large-scale parallel \textsc{C/C++} electronic structure code SPARC \cite{xu2021sparc, zhang2024sparc}.  In regards to the choice of exchange-correlation functional, we employ the Perdew-Zunger \cite{perdew1981self} (spin-unpolarized) and Perdew-Wang \cite{perdew1992accurate} (spin-polarized) variants  of LDA,  Perdew-Burke-Ernzerhof (PBE) \cite{perdew1996generalized} variant of GGA, rSCAN \cite{bartok2019regularized} variant of meta-GGA, and PBE0 \cite{adamo1999toward} variant of hybrid. In the PBE and PBE0 calculations, we employ ONCV \cite{hamann2013optimized} PBE pseudopotentials with nonlinear core correction from the SPMS set \cite{shojaei2023soft}. In the rSCAN calculations, we utilize ONCV pseudopotentials from the SG15 set \cite{schlipf2015optimization}.  In the LDA calculations, we use ONCV  pseudopotentials generated using the same parameters as those in the SPMS set. 

In calculations with local or semilocal exchange-correlation functionals, we solve the nonlinear eigenproblem (Eqn.~\ref{EL:Eigenproblem}) using a fixed-point iteration with respect to the density, referred to as the self-consistent field (SCF) iteration, which is accelerated by the restarted version of Periodic Pulay mixing \cite{banerjee2016periodic, pratapa2015restarted}. We use the density available in the pseudopotential file, partitioned into the spin-up and spin-down components based on the magnetization, as the starting guess. For hybrid exchange-correlation functionals, in addition to such an iteration, we employ an outer fixed-point iteration with respect to the exact-exchange operator. In this case, we first perform a PBE calculation and use the calculated orbitals and density as the starting guess. We solve both the eigenproblem and linear system using direct methods available within \textsc{MATLAB}. In the presentation of the results below, we will refer to the code implementing the spectral scheme as SPARC-atom, and we will use the maximum error in the eigenvalues as the metric for the accuracy,  with the corresponding error in total energy generally being one order of magnitude smaller. 

\begin{figure}[htbp]
            \centering
            \includegraphics[width=0.4\textwidth]{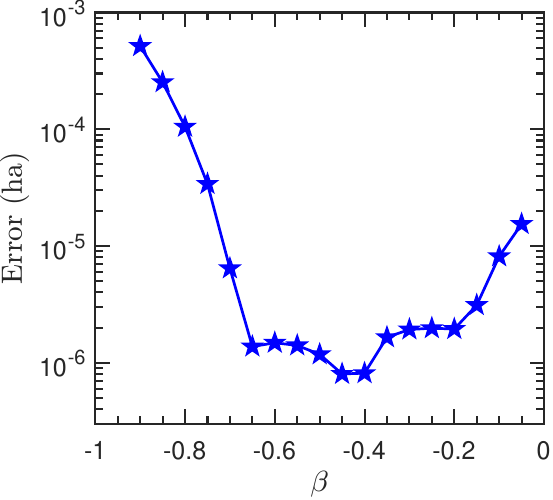}
            \caption{Error as a function of the exponential transformation parameter $\beta$   for the choice of PBE exchange-correlation functional, with $R=40$ bohr and $N_d=400$. The error is defined to be the maximum difference in eigenvalue over all the atoms in the SPMS set of pseudopotentials, the reference being the results obtained by the ONCVPSP  code.} 
            \label{fig:Mesh_parameter}
        \end{figure}

First, we optimize the parameter $\beta$ appearing in the transformation (Eqn.~\ref{Eqn:transformation})  for the choice of PBE exchange-correlation functional. In Fig.~\ref{fig:Mesh_parameter}, we present the variation in the error as a function of $\beta$, with the error defined as the maximum difference in eigenvalue over all the 69 atoms in the SPMS set of pseudopotentials, the reference being the ONCVPSP  code \cite{oncvpsp}. We observe that the error is relatively insensitive to $\beta$ in the range of -0.65 to -0.3, with -0.45 corresponding to the minimum. Indeed, we have found these results to be representative for the other choices of exchange-correlation functional, i.e., LDA, rSCAN, and hybrid. In view of this, we will  use $\beta=-0.45$ for the remainder of the simulations in this work.

\begin{figure}[htbp]
            \centering
            \subfloat{\includegraphics[width=0.4\textwidth]{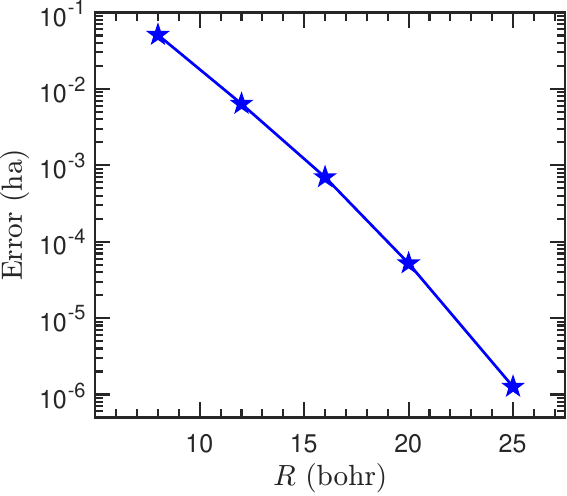}} \qquad \quad
            \subfloat{\includegraphics[width=0.4\textwidth]{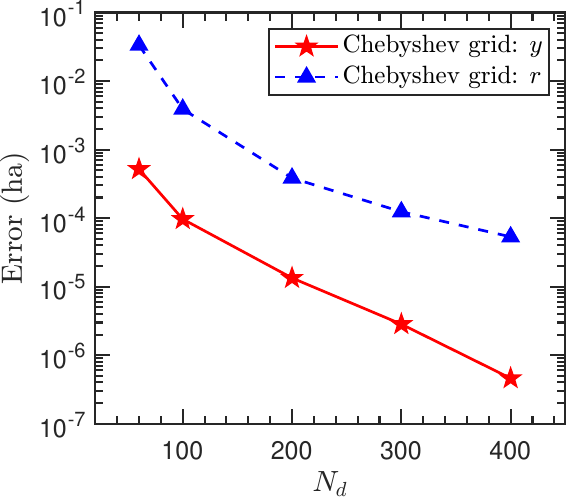}}
            \caption{Convergence with domain size $R$ and number of grid points $N_d$ for the PBE exchange-correlation functional \cite{perdew1996generalized}, with $\beta = -0.45$.  The error is defined to be the maximum difference in eigenvalue over all the atoms in the SPMS set of pseudopotentials, the reference being the results obtained by SPARC-atom for $R=40$ bohr and $N_d =2000$.}
            \label{fig:Convergence}
        \end{figure}

Next, we verify convergence of SPARC-atom with respect to the domain size $R$ and number of grid points $N_d$. We again consider all the atoms in the SPMS set of pseudopotentials and the  PBE exchange-correlation functional, the results so obtained  are presented in Fig.~\ref{fig:Convergence}. The error is defined to be the maximum difference in eigenvalue over all the atoms, the reference being the results obtained for  $R=40$ bohr and $N_d=2000$. We observe exponential convergence with respect to $R$, with $R=25$ bohr sufficient for achieving an accuracy of $10^{-6}$ ha. We also observe rapid convergence with respect to $N_d$, with $N_d \sim 400$ sufficient for achieving an accuracy of $10^{-6}$ ha. The transformed coordinate provides substantial advantages over the radial coordinate: $\sim 4$x reduction in the number of grid points for an accuracy of $10^{-4}$ ha, with an even larger reduction expected for an accuracy of $10^{-6}$ ha. The convergence with $N_d$ would have been even more rapid, if not for the reduced smoothness of the ONCV pseudopotentials, as demonstrated for all-electron DFT calculations in \ref{App:allelectron} and the radial Schr\"odinger equation in \ref{App:Schr}. Though we have found these results to be representative for the LDA and rSCAN exchange-correlation functionals, PBE0 requires $2-3$x larger number of grid points for achieving the same accuracy. This can be attributed to the exchange-operator (Eqn.~\ref{Vxx}), which requires a higher resolution relative to other parts of the Hamiltonian.

\begin{figure}[htbp]
            \centering
            \includegraphics[width=0.7\textwidth]{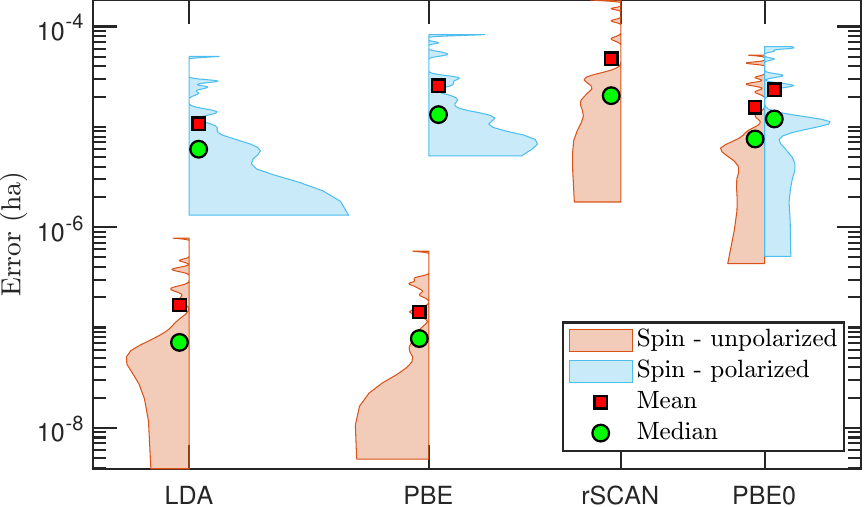}
            \caption{Error distribution for the choice of different exchange-correlation functionals, with $\beta=-0.45$, $R=40$ bohr, and $N_d=400$. The error is defined to be the maximum difference in eigenvalue over the He, N, O, Fe, Mn, Mo, and Cs atoms, the reference being the results obtained by the ONCVPSP  code for  spin-unpolarized LDA and PBE calculations, and the SPARC code for all other calculations.}
            \label{fig:violinXC}
        \end{figure}

Next, we verify the accuracy of SPARC-atom for different exchange-correlation functionals: LDA, PBE, rSCAN, and PBE0. We consider the following atoms: helium (He), nitrogen (N), oxygen (O), iron (Fe), manganese (Mn), molybdenum (Mo) and cesium (Cs), which have magnetizations of $0$, $3$, $2$, $4$, $5$, $6$, and  $1$, respectively. The results so obtained are summarized in Fig.~\ref{fig:violinXC}, with the data presented in Tables~\ref{tab:LDA}-\ref{tab:PBE0}. The reference for the spin-unpolarized LDA and PBE calculations is the ONCVPSP  code, which is restricted to such calculations, whereas the reference for other results is the  SPARC code in standard operation, where we employ 12th-order finite-differences, mesh-size of $0.15$ bohr, cubical domain with side of $60$ bohr, and Dirichlet boundary conditions. Due to SCF convergence issues in rSCAN  calculations for O and Cs  in SPARC, we instead employ periodic boundary conditions and a domain size of $30$ bohr. Indeed, the SCAN family of functionals is known to suffer from SCF convergence issues \cite{kingsbury2022performance, bartok2019regularized, lehtola2022many}, particularly for isolated systems in the presence of large vacuum. We have found that for the more popular r2SCAN  functional \cite{r2SCAN}, SPARC-atom faces SCF convergence issues at the stricter tolerances required for obtaining highly accurate results, which motivated the choice of  rSCAN here. The spin-polarized PBE0 calculations for Fe  failed to converge in SPARC, hence are not reported here. We observe from the results that there is excellent agreement between SPARC-atom and ONCVPSP, with all errors below $10^{-6}$ ha. We also observe that there is very good agreement between SPARC-atom and SPARC, with mean and median errors of $\sim 10^{-5}$ ha. Indeed, agreement increases further as the accuracy of SPARC calculations is improved by using finer meshes and larger vacuums. However, this becomes impractical to do for all systems, since for a mesh-size of 0.04 bohr, which is commensurate with the number of grid points required for an accuracy of $10^{-6}$ in SPARC-atom when using the Chebyshev grid in $r$ (Fig.~\ref{fig:Convergence}), there will be a total of more than 3.3 billion grid points in each SPARC calculation.

	   \begin{table}[htbp]
	   \small
	       \centering
              \caption{Comparison of eigenvalues for the choice of LDA exchange-correlation functional.}
	       \begin{tabular}{*{12}{c}}
                    \toprule
	             & $n$ & $l$ & $s$ & $g_{nl}$ & \multicolumn{3}{c}{Spin-unpolarized} & \multicolumn{3}{c}{Spin-polarized} \\
	            \midrule
                \multicolumn{5}{c}{} & SPARC-atom & ONCVPSP & Error & SPARC-atom & SPARC  & Error \\
                \midrule
                He & $1$ & $0$ & $\uparrow$ & $1$ & $-0.570229$ & $-0.570229$ & $7.8\eminus{7}$ & $-0.570289$ & $-0.570317$ & $2.8\eminus{5}$ \\
                &  &  & $\downarrow$ & $1$ & \multicolumn{3}{c}{} & $-0.570289$ & $-0.570317$ & $2.8\eminus{5}$ \\
                \midrule

                N & $2$ & $0$ & $\uparrow$ & $1$ & $-0.677041$ & $-0.677041$ & $1.9\eminus{7}$ & $-0.722837$ & $-0.722887$ & $4.9\eminus{5}$ \\
                &  &  & $\downarrow$ & $1$       & \multicolumn{3}{c}{} & $-0.559256$ & $-0.559307$ & $5.1\eminus{5}$ \\
                & $2$ & $1$ & $\uparrow$ & $3$   & $-0.266138$ & $-0.266138$ & $3.9\eminus{7}$ & $-0.308903$ & $-0.308875$ & $2.8\eminus{5}$ \\
                \midrule

                O & $2$ & $0$ & $\uparrow$ & $1$ & $-0.872914$ & $-0.872914$ & $4.6\eminus{7}$ & $-0.918374$ & $-0.918399$ & $2.5\eminus{5}$ \\
                &  &  & $\downarrow$ & $1$ & \multicolumn{3}{c}{} & $-0.801278$ & $-0.801303$ & $2.4\eminus{5}$ \\
                & $2$ & $1$ & $\uparrow$ & $3$   & $-0.338041$ & $-0.338041$ & $2.4\eminus{7}$ & $-0.381406$ & $-0.381420$ & $1.4\eminus{5}$ \\
                &  &  & $\downarrow$ & $1$ & \multicolumn{3}{c}{} & $-0.271167$ & $-0.271181$ & $1.4\eminus{5}$ \\
                \midrule
            
                Fe & $3$ & $0$ & $\uparrow$ & $1$ & $-3.436468$ & $-3.436468$ & $3.6\eminus{8}$ & $-3.471922$ & $-3.471927$ & $4.9\eminus{6}$ \\
                &  &  & $\downarrow$ & $1$ & \multicolumn{3}{c}{} & $-3.366835$ & $-3.366840$ & $4.9\eminus{6}$ \\
                & $3$ & $1$ & $\uparrow$ & $3$    & $-2.202261$ & $-2.202261$ & $4.8\eminus{8}$ & $-2.251541$ & $-2.251547$ & $6.3\eminus{6}$ \\
                &  &  & $\downarrow$ & $3$ & \multicolumn{3}{c}{} & $-2.118372$ & $-2.118378$ & $6.3\eminus{6}$ \\
                & $3$ & $2$ & $\uparrow$ & $5$    & $-0.284825$ & $-0.284825$ & $4.6\eminus{8}$ & $-0.335932$ & $-0.335938$ & $6.0\eminus{6}$ \\
                &  &  & $\downarrow$ & $1$ & \multicolumn{3}{c}{} & $-0.205288$ & $-0.205294$ & $5.9\eminus{6}$ \\
                & $4$ & $0$ & $\uparrow$ & $1$    & $-0.201710$ & $-0.201710$ & $4.5\eminus{8}$ & $-0.213154$ & $-0.213157$ & $2.6\eminus{6}$ \\
                &  &  & $\downarrow$ & $1$ & \multicolumn{3}{c}{} & $-0.185777$ & $-0.185779$ & $2.3\eminus{6}$ \\
                \midrule

                Mn & $3$ & $0$ & $\uparrow$ & $1$ & $-3.138528$ & $-3.138528$ & $9.3\eminus{8}$ & $-3.179547$ & $-3.179549$ & $2.2\eminus{6}$ \\
                &  &  & $\downarrow$ & $1$ & \multicolumn{3}{c}{} & $-3.024277$ & $-3.024279$ & $1.9\eminus{6}$ \\
                & $3$ & $1$ & $\uparrow$ & $3$    & $-2.002325$ & $-2.002325$ & $7.7\eminus{8}$ & $-2.054675$ & $-2.054677$ & $1.3\eminus{6}$ \\
                &  &  & $\downarrow$ & $3$ & \multicolumn{3}{c}{} & $-1.879342$ & $-1.879344$ & $1.1\eminus{6}$ \\
                & $3$ & $2$ & $\uparrow$ & $5$    & $-0.257509$ & $-0.257509$ & $3.7\eminus{7}$ & $-0.307410$ & $-0.307419$ & $9.9\eminus{6}$ \\
                & $4$ & $0$ & $\uparrow$ & $1$    & $-0.194387$ & $-0.194387$ & $2.6\eminus{10}$ & $-0.207673$ & $-0.207682$ & $7.7\eminus{6}$ \\
                &  &  & $\downarrow$ & $1$ & \multicolumn{3}{c}{} & $-0.171795$ & $-0.171803$ & $7.9\eminus{6}$ \\
                \midrule
                
                Mo & $4$ & $0$ & $\uparrow$ & $1$ & $-2.367515$ & $-2.367515$ & $2.2\eminus{8}$ & $-2.383101$ & $-2.383106$ & $5.0\eminus{6}$ \\
                &  &  & $\downarrow$ & $1$ & \multicolumn{3}{c}{} & $-2.290842$ & $-2.290848$ & $5.1\eminus{6}$ \\
                & $4$ & $1$ & $\uparrow$ & $3$    & $-1.418106$ & $-1.418106$ & $1.1\eminus{7}$ & $-1.441134$ & $-1.441137$ & $2.9\eminus{6}$ \\
                &  &  & $\downarrow$ & $3$ & \multicolumn{3}{c}{} & $-1.333709$ & $-1.333712$ & $2.9\eminus{6}$ \\
                & $4$ & $2$ & $\uparrow$ & $5$    & $-0.144627$ & $-0.144628$ & $2.4\eminus{7}$ & $-0.172029$ & $-0.172042$ & $1.4\eminus{6}$ \\
                & $5$ & $0$ & $\uparrow$ & $1$    & $-0.159104$ & $-0.159104$ & $5.7\eminus{8}$ & $-0.177067$ & $-0.177069$ & $2.6\eminus{6}$ \\
                \bottomrule

                Cs & $5$ & $0$ & $\uparrow$ & $1$ & $-0.988405$ & $-0.988405$ & $5.0\eminus{8}$ & $-0.987361$ & $-0.987363$ & $1.7\eminus{6}$ \\
                &  &  & $\downarrow$ & $1$  & \multicolumn{3}{c}{} & $-0.986088$ & $-0.986089$ & $1.6\eminus{6}$ \\
                & $5$ & $1$ & $\uparrow$ & $3$    & $-0.500051$ & $-0.500051$ & $6.4\eminus{8}$ & $-0.499546$ & $-0.499548$ & $1.8\eminus{6}$ \\
                &  &  & $\downarrow$ & $3$ & \multicolumn{3}{c}{} & $-0.496973$ & $-0.496975$ & $1.9\eminus{6}$ \\
                & $6$ & $0$ & $\uparrow$ & $1$    & $-0.081848$ & $-0.081848$ & $2.2\eminus{8}$ & $-0.087726$ & $-0.087728$ & $1.6\eminus{6}$ \\
                \midrule
                                
	       \end{tabular}
	       
	       \label{tab:LDA}
	   \end{table}

          \begin{table}[htbp]
            \small
	       \centering
              \caption{Comparison of the eigenvalues for the choice of PBE exchange-correlation functional.}
	       \begin{tabular}{*{12}{c}}
                    \toprule
	             & $n$ & $l$ & $s$ & $g_{nl}$ & \multicolumn{3}{c}{Spin-unpolarized} & \multicolumn{3}{c}{Spin-polarized} \\
	            \midrule
                               \multicolumn{5}{c}{} & SPARC-atom & ONCVPSP  & Error & SPARC-atom & SPARC  & Error \\
                \midrule
                He & $1$ & $0$ & $\uparrow$ & $1$ & $-0.579311$ & $-0.579311$ & $1.4\eminus7$ & $-0.579311$ & $-0.579342$ & $3.1\eminus5$ \\
                &  &  & $\downarrow$ & $1$ & \multicolumn{3}{c}{} & $-0.579311$ & $-0.579342$ & $3.1\eminus5$ \\
                \midrule

                N & $2$ & $0$ & $\uparrow$ & $1$ & $-0.682914$ & $-0.682914$ & $2.7\eminus7$ & $-0.732055$ & $-0.732106$ & $5.1\eminus5$ \\
                &  &  & $\downarrow$ & $1$       & \multicolumn{3}{c}{} & $-0.553662$ & $-0.553715$ & $5.3\eminus5$ \\
                & $2$ & $1$ & $\uparrow$ & $3$   & $-0.260551$ & $-0.260551$ & $3.2\eminus7$ & $-0.306694$ & $-0.306667$ & $2.6\eminus5$ \\
                \midrule

                O & $2$ & $0$ & $\uparrow$ & $1$ & $-0.880576$ & $-0.880576$ & $5.8\eminus7$ & $-0.928616$ & $-0.928646$ & $2.9\eminus5$ \\
                &  &  & $\downarrow$ & $1$ & \multicolumn{3}{c}{} & $-0.803758$ & $-0.803786$ & $2.8\eminus5$ \\
                & $2$ & $1$ & $\uparrow$ & $3$   & $-0.331872$ & $-0.331872$ & $1.5\eminus7$ & $-0.378221$ & $-0.378238$ & $1.7\eminus5$ \\
                &  &  & $\downarrow$ & $1$ & \multicolumn{3}{c}{} & $-0.259744$ & $-0.259757$ & $1.3\eminus5$ \\
                \midrule
            
                Fe & $3$ & $0$ & $\uparrow$ & $1$ & $-3.455077$ & $-3.455077$ & $1.8\eminus8$ & $-3.487629$ & $-3.487636$ & $6.6\eminus6$ \\
                &  &  & $\downarrow$ & $1$ & \multicolumn{3}{c}{} & $-3.386986$ & $-3.386992$ & $6.5\eminus6$ \\
                & $3$ & $1$ & $\uparrow$ & $3$    & $-2.206535$ & $-2.206535$ & $2.9\eminus8$ & $-2.257080$ & $-2.257088$ & $7.9\eminus6$ \\
                &  &  & $\downarrow$ & $3$ & \multicolumn{3}{c}{} & $-2.118574$ & $-2.118582$ & $8.2\eminus6$ \\
                & $3$ & $2$ & $\uparrow$ & $5$    & $-0.275801$ & $-0.275801$ & $9.4\eminus9$ & $-0.331647$ & $-0.331656$ & $8.5\eminus6$ \\
                &  &  & $\downarrow$ & $1$ & \multicolumn{3}{c}{} & $-0.187754$ & $-0.187763$ & $8.9\eminus6$ \\
                & $4$ & $0$ & $\uparrow$ & $1$    & $-0.194482$ & $-0.194482$ & $5.9\eminus8$ & $-0.205837$ & $-0.205845$ & $7.3\eminus6$ \\
                &  &  & $\downarrow$ & $1$ & \multicolumn{3}{c}{} & $-0.178800$ & $-0.178808$ & $7.3\eminus6$ \\
                \midrule

                Mn & $3$ & $0$ & $\uparrow$ & $1$ & $-3.156010$ & $-3.156010$ & $1.6\eminus8$ & $-3.195297$ & $-3.195310$ & $1.3\eminus5$ \\
                &  &  & $\downarrow$ & $1$ & \multicolumn{3}{c}{} & $-3.036206$ & $-3.036218$ & $1.3\eminus5$ \\
                & $3$ & $1$ & $\uparrow$ & $3$    & $-2.006265$ & $-2.006265$ & $2.0\eminus9$ & $-2.060678$ & $-2.060689$ & $1.2\eminus5$ \\
                &  &  & $\downarrow$ & $3$ & \multicolumn{3}{c}{} & $-1.872649$ & $-1.872660$ & $1.1\eminus5$ \\
                & $3$ & $2$ & $\uparrow$ & $5$    & $-0.248834$ & $-0.248834$ & $2.9\eminus7$ & $-0.301704$ & $-0.301723$ & $1.9\eminus5$ \\
                & $4$ & $0$ & $\uparrow$ & $1$    & $-0.187687$ & $-0.187687$ & $1.8\eminus8$ & $-0.200252$ & $-0.200265$ & $1.2\eminus5$ \\
                &  &  & $\downarrow$ & $1$ & \multicolumn{3}{c}{} & $-0.165445$ & $-0.165457$ & $1.2\eminus5$ \\
                \midrule
                
                Mo & $4$ & $0$ & $\uparrow$ & $1$ & $-2.364713$ & $-2.364713$ & $1.4\eminus8$ & $-2.377416$ & $-2.377423$ & $7.5\eminus6$ \\
                &  &  & $\downarrow$ & $1$ & \multicolumn{3}{c}{} & $-2.284147$ & $-2.284155$ & $7.8\eminus6$ \\
                & $4$ & $1$ & $\uparrow$ & $3$    & $-1.414321$ & $-1.414321$ & $6.1\eminus8$ & $-1.436952$ & $-1.436958$ & $5.6\eminus6$ \\
                &  &  & $\downarrow$ & $3$ & \multicolumn{3}{c}{} & $-1.322545$ & $-1.322551$ & $5.6\eminus6$ \\
                & $4$ & $2$ & $\uparrow$ & $5$    & $-0.137922$ & $-0.137922$ & $2.1\eminus7$ & $-0.168243$ & $-0.168258$ & $1.5\eminus5$ \\
                & $5$ & $0$ & $\uparrow$ & $1$    & $-0.150182$ & $-0.150182$ & $9.4\eminus8$ & $-0.169038$ & $-0.169042$ & $4.7\eminus6$ \\
                \bottomrule

                Cs & $5$ & $0$ & $\uparrow$ & $1$ & $-0.982389$ & $-0.982389$ & $2.7\eminus7$ & $-0.981385$ & $-0.981467$ & $8.2\eminus5$ \\
                &  &  & $\downarrow$ & $1$  & \multicolumn{3}{c}{} & $-0.979828$ & $-0.979912$ & $8.3\eminus5$ \\
                & $5$ & $1$ & $\uparrow$ & $3$    & $-0.496788$ & $-0.496788$ & $2.4\eminus7$ & $-0.496592$ & $-0.496673$ & $8.1\eminus5$ \\
                &  &  & $\downarrow$ & $3$ & \multicolumn{3}{c}{} & $-0.493419$ & $-0.493502$ & $8.3\eminus5$ \\
                & $6$ & $0$ & $\uparrow$ & $1$    & $-0.076669$ & $-0.076669$ & $3.3\eminus8$ & $-0.083330$ & $-0.083386$ & $5.6\eminus5$ \\
                \midrule
                                
	       \end{tabular}
	       
	       \label{tab:GGA}
	   \end{table}

	
        \begin{table}[htbp]
          \small
	       \centering
              \caption{Comparison of the eigenvalues for the choice of rSCAN exchange-correlation functional.}
	       \begin{tabular}{*{9}{c}}
             \toprule
	        & $n$ & $l$ & $g_{nl}$ & \multicolumn{3}{c}{Spin-unpolarized}\\
	        \midrule
                \multicolumn{4}{c}{} & SPARC-atom & SPARC & Error \\
                \midrule
                
                He & $1$ & $0$ & $2$ & $-0.605361$ & $-0.605362$ & $8.6\eminus7$  \\
                
                \midrule

                N & $2$ & $0$ & $2$ & $-0.729362$ & $-0.729287$ & $7.5\eminus5$ \\
                
                & $2$ & $1$ & $3$   & $-0.265639$ & $-0.265650$ & $1.1\eminus5$ \\
                \midrule

                O$^\dag$ & $2$ & $0$ & $2$ & $-0.940105$ & $-0.940097$ & $8.3\eminus6$  \\
                
                & $2$ & $1$ & $4$   & $-0.343355$ & $-0.343363$ & $8.3\eminus6$  \\
                
                \midrule
            
                Fe & $3$ & $0$ & $2$ & $-3.585987$ & $-3.585957$ & $2.9\eminus5$  \\
                
                & $3$ & $1$ & $6$    & $-2.293532$ & $-2.293512$ & $1.9\eminus5$  \\
                
                & $3$ & $2$ & $6$    & $-0.281911$ & $-0.281907$ & $3.9\eminus6$  \\
                
                & $4$ & $0$ & $2$    & $-0.203008$ & $-0.203000$ & $7.6\eminus6$  \\
                
                \midrule

                Mn & $3$ & $0$ & $2$ & $-3.279161$ & $-3.279139$ & $2.1\eminus5$  \\
                
                & $3$ & $1$ & $6$    & $-2.085219$ & $-2.085205$ & $1.4\eminus5$  \\
                
                & $3$ & $2$ & $5$    & $-0.251825$ & $-0.251818$ & $7.1\eminus6$  \\
                
                & $4$ & $0$ & $2$    & $-0.195917$ & $-0.195909$ & $6.8\eminus6$  \\
                
                \midrule
                
                Mo & $4$ & $0$ & $2$ & $-2.456659$ & $-2.456676$ & $1.6\eminus5$  \\
                
                & $4$ & $1$ & $6$    & $-1.469806$ & $-1.469772$ & $3.4\eminus5$  \\
                
                & $4$ & $2$ & $5$    & $-0.139040$ & $-0.139012$ & $2.8\eminus5$  \\
                
                & $5$ & $0$ & $1$    & $-0.154164$ & $-0.154146$ & $1.8\eminus5$  \\
                \bottomrule

                Cs$^\dag$ & $5$ & $0$ & $2$ & $-1.023124$ & $-1.023145$ & $2.1\eminus5$  \\
                
                & $5$ & $1$ & $6$    & $-0.518869$ & $-0.518889$ & $1.9\eminus5$  \\
                
                & $6$ & $0$ & $1$    & $-0.076437$ & $-0.076396$ & $4.1\eminus5$  \\
                \midrule                
	       \end{tabular}
	       \label{tab:RSCAN}
	   \end{table}

        \begin{table}[htbp]
          \small
	       \centering
              \caption{Comparison of the eigenvalues for the choice of PBE0 exchange-correlation functional.}
	       \begin{tabular}{*{12}{c}}
                    \toprule
	             & $n$ & $l$ & $s$ & $g_{nl}$ & \multicolumn{3}{c}{Spin-unpolarized} & \multicolumn{3}{c}{Spin-polarized} \\
	            \midrule
                \multicolumn{5}{c}{} & SPARC-atom & SPARC  & Error & SPARC-atom & SPARC  & Error \\
                \midrule
                He & $1$ & $0$ & $\uparrow$ & $1$ & $-0.668804$ & $-0.668831$ & $2.7\eminus5$ & $-0.668804$ & $-0.668831$ & $2.7\eminus5$ \\
                &  &  & $\downarrow$ & $1$ & \multicolumn{3}{c}{} & $-0.668804$ & $-0.668831$ & $2.7\eminus5$ \\
                \midrule

                N & $2$ & $0$ & $\uparrow$ & $1$ & $-0.781194$ & $-0.781246$ & $5.2\eminus5$ & $-0.849158$ & $-0.849205$ & $4.8\eminus5$ \\
                &  &  & $\downarrow$ & $1$       & \multicolumn{3}{c}{} & $-0.610421$ & $-0.610478$ & $5.6\eminus5$ \\
                & $2$ & $1$ & $\uparrow$ & $3$   & $-0.268919$ & $-0.268892$ & $2.6\eminus5$ & $-0.376158$ & $-0.376126$ & $3.2\eminus5$ \\
                \midrule

                O & $2$ & $0$ & $\uparrow$ & $1$ & $-0.999299$ & $-0.999321$ & $2.2\eminus5$ & $-1.073173$ & $-1.073195$ & $2.2\eminus5$ \\
                &  &  & $\downarrow$ & $1$ & \multicolumn{3}{c}{} & $-0.889919$ & $-0.889945$ & $2.6\eminus5$ \\
                & $2$ & $1$ & $\uparrow$ & $3$   & $-0.364142$ & $-0.364151$ & $9.1\eminus6$ & $-0.468697$ & $-0.468707$ & $9.7\eminus6$ \\
                &  &  & $\downarrow$ & $1$ & \multicolumn{3}{c}{} & $-0.232362$ & $-0.232374$ & $1.1\eminus5$ \\
                \midrule
            
                Fe & $3$ & $0$ & $\uparrow$ & $1$ & $-3.665568$ & $-3.665562$ & $5.4\eminus6$ & $-3.708520$ & --- & --- \\
                &  &  & $\downarrow$ & $1$ & \multicolumn{3}{c}{} & $-3.570388$ & --- & --- \\
                & $3$ & $1$ & $\uparrow$ & $3$    & $-2.364003$ & $-2.363997$ & $6.8\eminus6$ & $-2.439943$ & --- & --- \\
                &  &  & $\downarrow$ & $3$ & \multicolumn{3}{c}{} & $-2.238089$ & --- & --- \\
                & $3$ & $2$ & $\uparrow$ & $5$    & $-0.258819$ & $-0.258832$ & $1.3\eminus5$ & $-0.424505$ & --- & --- \\
                &  &  & $\downarrow$ & $1$ & \multicolumn{3}{c}{} & $-0.137454$ & --- & --- \\
                & $4$ & $0$ & $\uparrow$ & $1$    & $-0.222093$ & $-0.222093$ & $1.7\eminus7$ & $-0.233262$ & --- & --- \\
                &  &  & $\downarrow$ & $1$ & \multicolumn{3}{c}{} & $-0.205489$ & --- & --- \\
                \midrule

                Mn & $3$ & $0$ & $\uparrow$ & $1$ & $-3.363101$ & $-3.363104$ & $2.4\eminus6$ & $-3.402252$ & $-3.402250$ & $1.5\eminus6$ \\
                &  &  & $\downarrow$ & $1$ & \multicolumn{3}{c}{} & $-3.191481$ & $-3.191494$ & $1.3\eminus5$ \\
                & $3$ & $1$ & $\uparrow$ & $3$    & $-2.162655$ & $-2.162657$ & $2.4\eminus6$ & $-2.234814$ & $-2.234808$ & $5.2\eminus6$ \\
                &  &  & $\downarrow$ & $3$ & \multicolumn{3}{c}{} & $-1.968028$ & $-1.968040$ & $1.2\eminus5$ \\
                & $3$ & $2$ & $\uparrow$ & $5$    & $-0.258819$ & $-0.258832$ & $1.3\eminus5$ & $-0.387727$ & $-0.387734$ & $6.2\eminus6$ \\
                & $4$ & $0$ & $\uparrow$ & $1$    & $-0.214786$ & $-0.214794$ & $8.4\eminus6$ & $-0.224147$ & $-0.224158$ & $1.1\eminus5$ \\
                &  &  & $\downarrow$ & $1$ & \multicolumn{3}{c}{} & $-0.188822$ & $-0.188832$ & $1.0\eminus5$ \\
                \midrule
                
                Mo & $4$ & $0$ & $\uparrow$ & $1$ & $-2.526864$ & $-2.526861$ & $3.7\eminus6$ & $-2.528792$ & $-2.528796$ & $3.6\eminus6$ \\
                &  &  & $\downarrow$ & $1$ & \multicolumn{3}{c}{} & $-2.407219$ & $-2.407231$ & $1.1\eminus5$ \\
                & $4$ & $1$ & $\uparrow$ & $3$    & $-1.531114$ & $-1.531107$ & $6.4\eminus6$ & $-1.553601$ & $-1.553601$ & $1.9\eminus7$ \\
                &  &  & $\downarrow$ & $3$ & \multicolumn{3}{c}{} & $-1.394790$ & $-1.394798$ & $8.5\eminus6$ \\
                & $4$ & $2$ & $\uparrow$ & $5$    & $-0.144338$ & $-0.144344$ & $5.9\eminus6$ & $-0.216322$ & $-0.216333$ & $1.1\eminus5$ \\
                & $5$ & $0$ & $\uparrow$ & $1$    & $-0.147807$ & $-0.147808$ & $6.6\eminus7$ & $-0.188873$ & $-0.188877$ & $3.5\eminus6$ \\
                \bottomrule

                Cs & $5$ & $0$ & $\uparrow$ & $1$ & $-1.066612$ & $-1.066656$ & $4.3\eminus5$ & $-1.062389$ & $-1.062449$ & $5.9\eminus5$ \\
                &  &  & $\downarrow$ & $1$  & \multicolumn{3}{c}{} & $-1.060152$ & $-1.060213$ & $6.1\eminus5$ \\
                & $5$ & $1$ & $\uparrow$ & $3$    & $-0.551294$ & $-0.551337$ & $4.3\eminus5$ & $-0.547484$ & $-0.547545$ & $6.1\eminus5$ \\
                &  &  & $\downarrow$ & $3$ & \multicolumn{3}{c}{} & $-0.544656$ & $-0.544719$ & $6.3\eminus5$ \\
                & $6$ & $0$ & $\uparrow$ & $1$    & $-0.076059$ & $-0.076089$ & $3.1\eminus5$ & $-0.097692$ & $-0.097725$ & $3.3\eminus5$ \\
                \midrule
                                
	       \end{tabular}
	       
	       \label{tab:PBE0}
	   \end{table}

Finally, we discuss the computational aspects of atomic structure calculations in SPARC-atom. The key computational kernels are the the linear solver, which is invoked once each SCF iteration, and the dense eigensolver, which is invoked a maximum of four times in each SCF iteration, one for each of the occupied azimuthal quantum numbers. The eigensolver is the dominant cost, taking $>92$\% and $>96$\% of the total time for $N_d=200$  and $N_d=400$, respectively. The overall cost of atomic structure calculations scales cubically with the number of grid points, i.e., $\mathcal{O}(N_d^3)$.  Given that the number of grid points and the prefactor are both small in the spectral scheme, the calculations are found to be particularly efficient. For instance, spin-unpolarized self-consistent DFT calculations of Mn, which require $N_d=200$, $200$, $250$, and $400$ for accuracy of $10^{-6}$ ha in LDA, PBE, rSCAN, and PBE0 calculations, take $0.36$, $0.38$, $3.82$, and $10.8$ seconds, respectively.  The significantly larger time taken by rSCAN and PBE0, in addition to the $\mathcal{O}(N_d^3)$ scaling, is a consequence of the number of SCF iterations being a factor of $\sim 6$x larger and $\sim 7$x larger, respectively. Indeed, these timings will be further reduced with an efficient  implementation in a lower level programming language such as \textsc{C/C++}.

	
	\section{Concluding remarks} \label{conclusions}
       In this work, we have developed a spectral scheme for atomic structure calculations in pseudopotential Kohn-Sham DFT. In particular, following an exponential transformation of the radial coordinates, we have employed global polynomial interpolation on a Chebyshev grid, with the associated Chebyshev differentiation matrix and Clenshaw-Curtis quadrature used to approximate derivative operators and integrals, respectively.   We have demonstrated the accuracy and efficiency of the scheme through spin-polarized and unpolarized calculations for representative atoms, while considering the LDA, PBE, rSCAN, and PBE0 exchange-correlation functionals, choices that are representative of the different levels of complexity in the DFT formalism. In particular, considering ONCV pseudopotentials spanning the periodic table from atomic number $Z = 1$ to $83$,  we have found that the spectral scheme achieves an accuracy of 1 microhartree in the eigenvalues  for $\mathcal{O}$(200) grid points, a number that is smaller still for smoother potentials.

The current implementation has been developed in M-SPARC, the \textsc{MATLAB} version of the large-scale parallel \textsc{C/C++} code SPARC. Opportunities for future work include developing an implementation in the SPARC code and extending the formalism to higher levels of theory such the random phase approximation and quantum Monte Carlo.

    

	\section*{Acknowledgments}
	The authors gratefully acknowledge the support of the U.S. Department of Energy, Office of Science under grant DE-SC0023445. This work was performed in part under the auspices of the U.S. Department of Energy by Lawrence Livermore National Laboratory under Contract DE-AC52-07NA27344. This research was also supported by the supercomputing infrastructure provided by Partnership for an Advanced Computing Environment (PACE) through its Hive (U.S. National Science Foundation through grant MRI1828187) and Phoenix clusters at Georgia Institute of Technology, Atlanta, Georgia.


\appendix

\section{All-electron DFT with LDA exchange-correlation} \label{App:allelectron}
\begin{figure}[htbp]
            \centering
            \subfloat{\includegraphics[width=0.4\textwidth]{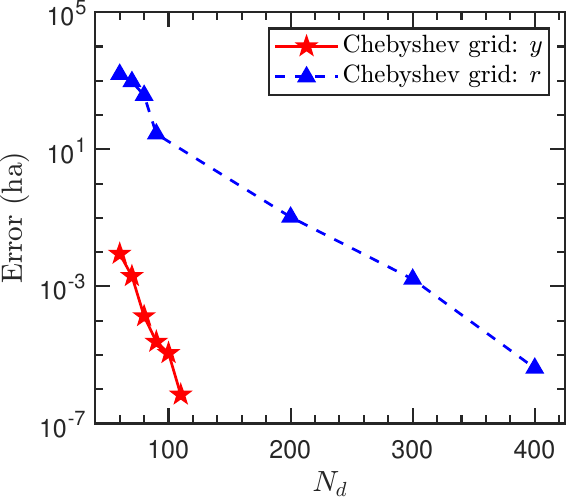}} \qquad \quad 
            \subfloat{\includegraphics[width=0.4\textwidth]{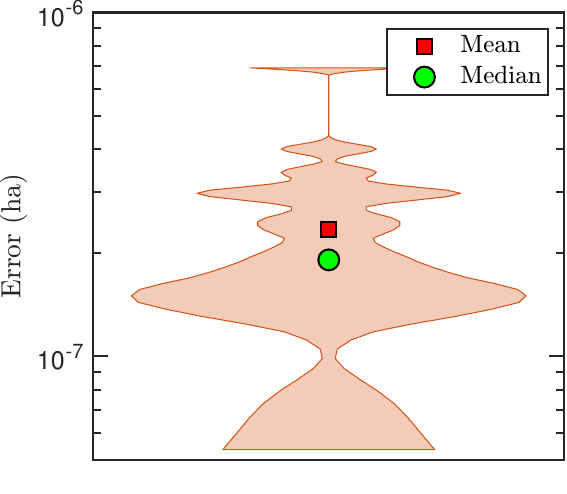}}
            \caption{Accuracy of the spectral scheme for all-electron DFT with LDA exchange-correlation for $Z=92$. The error is defined to be the maximum difference in eigenvalue, the reference being NIST. Left: Convergence of the eigenvalues with respect to number of grid points $N_d$.  Right: Violin plot for the distribution of the error in the eigenvalues for $N_d=110$.}
            \label{fig:All-electronDFT}
        \end{figure}

We now demonstrate the accuracy and efficiency of the spectral scheme for all-electron Kohn-Sham DFT calculations with the LDA exchange-correlation:
        \begin{equation}\label{SchroTilde}
           \left(  -\frac{1}{2} \frac{d^2}{dr^2}  + V_{xc}^{LDA} + \frac{\widetilde{V}_H}{r} -\frac{Z}{r}  + \frac{l(l+1)}{2r^2} \right)  \widetilde{R}_{nl} = \lambda_{nl} \widetilde{R}_{nl} \,,
        \end{equation}
obtained by setting  $V_{loc} = -Z/r$ and $V_{nl}=0$   in   Eqn.~\ref{EL:Eigenproblem}. In Fig.~\ref{fig:All-electronDFT}, we compare the results so obtained for $Z=92$ with the NIST values  \cite{nist_data}, both corresponding to the Vosko-Wilk-Nusair (VWN) \cite{LDA_VWN} variant of the LDA.  It is clear that there is rapid convergence in the eigenvalues, with an accuracy of better than $10^{-6}$ ha achieved with only $N_d \sim 110$ grid points in the transformed coordinates, which translates to a wall time of $\sim$0.68 s, the calculation requiring 18 SCF iterations for a Gaussian starting guess. The requirement of substantially smaller number of grid points relative to pseudopotential DFT  can be attributed to reduced smoothness of the Kohn-Sham problem within the ONCV pseudopotential approximation. These results suggest that the spectral scheme can be particularly efficient for smooth pseudopotentials \cite{goedecker1996separable, hartwigsen1998relativistic}. As in pseudopotential DFT, the transformed coordinate provide substantial advantages over the radial coordinate, with a $\sim 4$x reduction in the number of grid points for an accuracy of $10^{-6}$ ha. 

\section{Radial Schr\"odinger equation} \label{App:Schr}
\begin{figure}[htbp]
            \centering
            \subfloat{\includegraphics[width=0.4\textwidth]{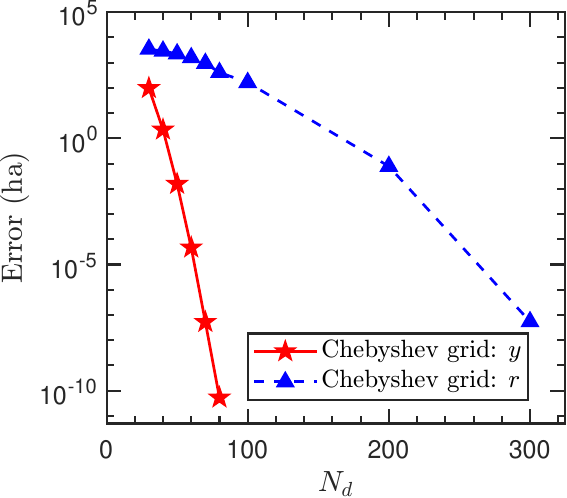}} \qquad \quad 
            \subfloat{\includegraphics[width=0.4\textwidth]{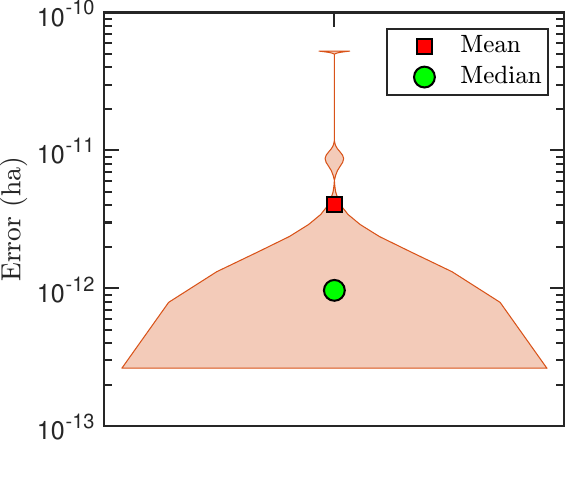}}
            \caption{Accuracy of the spectral scheme for solution of the radial Schr\"odinger equation for $Z=92$. The error is defined with respect to the exact theoretical values. Left: Convergence of the eigenvalues with respect to number of grid points $N_d$.  Right: Violin plot for the distribution of the error in the eigenvalues for $N_d=80$.}
            \label{fig:Schrodinger}
        \end{figure}

We now demonstrate the accuracy and efficiency of the spectral scheme for the all-electron radial Schr\"odinger equation \cite{vcertik2024high, vcertik2013dftatom}:
        \begin{equation}\label{SchroTilde}
           \left(  -\frac{1}{2} \frac{d^2}{dr^2}  -\frac{Z}{r}  + \frac{l(l+1)}{2r^2} \right)  \widetilde{R}_{nl} = \lambda_{nl} \widetilde{R}_{nl} \,,
        \end{equation}
which represents the linearized version of the radial Kohn-Sham DFT equation (Eqn.~\ref{EL:Eigenproblem}), obtained by setting $V_{loc} = -Z/r$ and $V_{xc}=V_{nl}=\widetilde{V}_H=0$. In Fig.~\ref{fig:Schrodinger}, we compare the results so obtained for $Z=92$ with the exact theoretical result. It is clear that there is extremely rapid convergence in the eigenvalues, with an accuracy of better than $10^{-10}$ ha achieved for only $N_d \sim 80$ grid points in the transformed coordinates. In particular, $N_d \sim 65$ grid points are sufficient to achieve an accuracy of $10^{-6}$ ha, which translates to a wall time of $\sim$0.018 s. The requirement of substantially smaller number of grid points relative to DFT  can be attributed to reduced smoothness of the DFT problem, particularly within the pseudopotential approximation. As in  DFT, the transformed coordinate provide substantial advantages over the radial coordinate, with a $\sim 5$x reduction in the number of grid points for an accuracy of $10^{-6}$ ha.

	\bibliographystyle{unsrt}


\end{document}